# The optical design of the six-meter CCAT-prime and Simons Observatory telescopes

Stephen C. Parshley[*a], Michael D. Niemack[b], Richard Hills[c], Simon R. Dicker[d], Rolando Dünner[e], Jens Erler[f], Patricio A. Gallardo[b], Jon E. Gudmundsson[g], Terry Herter[a], Brian J. Koopman[b], Michele Limon[d], Frederick T. Matsuda[h], Philip Mauskopf[i], Dominik A. Riechers[a], Gordon J. Stacey[a], Eve M. Vavagiakis[b]

[a]Department of Astronomy, Cornell University, Ithaca, NY 14853, USA
[b]Department of Physics, Cornell University, Ithaca, NY 14853, USA
[c]Cavendish Laboratory, University of Cambridge, Cambridge CB3 0HE, UK
[d]Department of Physics and Astronomy, University of Pennsylvania, Philadelphia, PA, USA
[e]Instituto de Astrofísica y Centro de Astroingeniería, Facultad de Física, Pontificia Universidad Católica de Chile, 7820436 Macul, Santiago, Chile
[f]Argelander-Institute für Astronomie, Rheinische Friedrich-Wilhelms Universität Bonn, 53121 Bonn, Germany
[g]The Oskar Klein Centre for Cosmoparticle Physics, Department of Physics, Stockholm University, Stockholm, Sweden
[h]Kavli Institute for the Physics and Mathematics of the Universe (WPI), The University of Tokyo Institutes for Advanced Study, University of Tokyo, Kashiwa Chiba 277-8583, Japan
[i]School of Earth and Space Exploration and Department of Physics, Arizona State University, Tempe, AZ 85287, USA

**ABSTRACT**

A common optical design for a coma-corrected, 6-meter aperture, crossed-Dragone telescope has been adopted for the CCAT-prime telescope of CCAT Observatory, Inc., and for the Large Aperture Telescope of the Simons Observatory. Both are to be built in the high altitude Atacama Desert in Chile for submillimeter and millimeter wavelength observations, respectively. The design delivers a high throughput, relatively flat focal plane, with a field of view 7.8 degrees in diameter for 3 mm wavelengths, and the ability to illuminate >100k diffraction-limited beams for < 1 mm wavelengths. The optics consist of offset reflecting primary and secondary surfaces arranged in such a way as to satisfy the Mizuguchi-Dragone criterion, suppressing first-order astigmatism and maintaining high polarization purity. The surface shapes are perturbed from their standard conic forms in order to correct coma aberrations. We discuss the optical design, performance, and tolerancing sensitivity. More information about CCAT-prime can be found at ccatobservatory.org and about Simons Observatory at simonsobservatory.org.

## 1 INTRODUCTION

The CCAT-prime telescope and the Simons Observatory (SO) Large Aperture Telescope (LAT) share a common crossed-Dragone[1] optical design, with a 6-meter aperture and a fairly flat focal plane. The optical throughput is high, with each telescope being able to field ~10 times more detectors on-sky than current observatories[2]. SO LAT will be dedicated to millimeter (mm) wavelength observations of the cosmic microwave background (CMB). CCAT-prime will be a multi-instrument platform operating in the millimeter and submillimeter (submm) wavelength regimes, exploring the epoch of reionization, large-scale filaments and diffuse atomic gas, giant molecular clouds and star-forming regions, dark energy, distant galaxies and galaxy clusters, as well as the CMB[3]. Water vapor in the Earth's atmosphere absorbs the wavelengths to be studied, so both telescopes will be located in the high altitude Atacama region in Chile, near the Atacama Large Millimeter Array (ALMA)[4]. The CCAT site is roughly 400 m above the SO LAT site and offers significant gains[5] in the telluric window transmission, and routine access to the short submm bands (< 350 μm).

[*] scp8@cornell.edu; phone 1 (607) 255-4806; www.ccatobservatory.org

This paper presents the optical design, performance, and tolerancing sensitivity of the CCAT-prime telescope and SO LAT. Additional details about CCAT-prime and Simons Observatory can be found in other manuscripts presented at this conference. For details on the mechanical telescope design see Parshley et al. 2018[6]. A CCAT first-light camera design is discussed in Vavagiakis et al. 2018[7]. An SO science book is in development, an overview of SO instrumentation is presented in Galitzki et al. 2018[8], optical systematics for this telescope design are described in Gallardo et al. 2018[9], cryogenic optics designs are in Dicker et al. 2018[10], and design analyses for the SO LAT receiver are presented in Zhu et al. 2018[11], Orlowski-Scherer et al. 2018[12], and Coppi et al. 2018[13].

## 2 OPTICAL DESIGN

Low sidelobes and optical loading are an essential requirement for advanced CMB studies, including CMB polarization studies, which can readily be achieved if the telescope has a clear aperture, i.e. that there is no blockage caused by the secondary mirror or the instruments[14, 15, 16, 17]. This typically means that off-axis mirrors are used[18]. We adopted the crossed-Dragone configuration in which the first-order aberration and cross-polarization caused by the tilt of the primary are cancelled by the choice of the tilt of the secondary. The optics for CCAT-prime are shown in Figure 1. The SO LAT optics are identical, but the subordinate instrument spaces are reallocated for science support equipment instead of additional focal plane instrumentation via reimaging optics as in CCAT-prime. The design is based on the classical designs studied in Niemack[2] and modified to correct for coma as detailed below. The starting point was a classical design satisfying the Mizuguchi-Dragone criterion. This results in an unobstructed, high-throughput telescope with low cross-polarization and a minimally curved focal plane, suitable for large cameras.

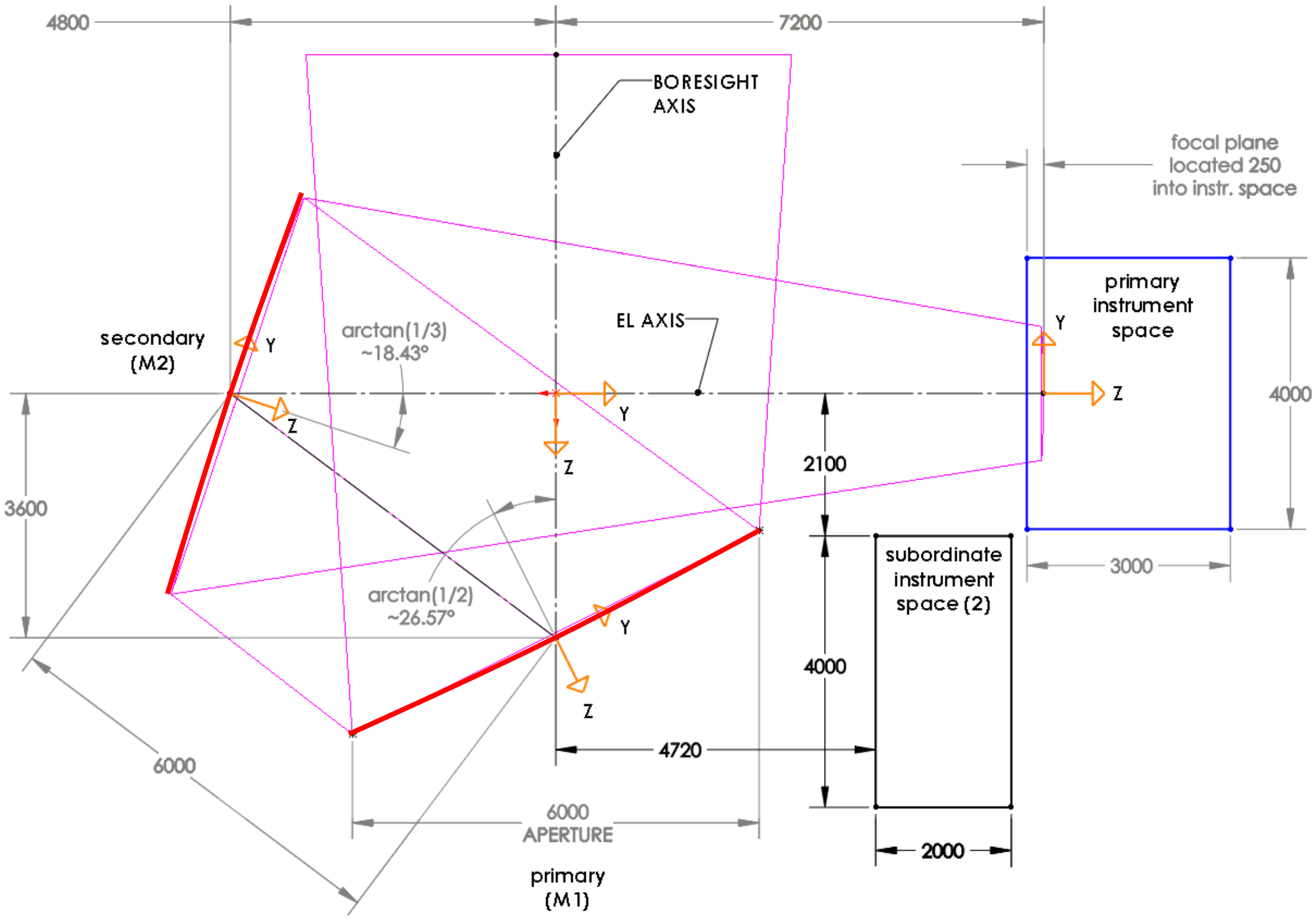

Figure 1. 2D layout slice showing the modified crossed-Dragone optical design for CCAT-prime. The optical beam is magenta and mirror sections are red. All linear dimensions are in millimeters, and numbers without decimal points are exact. Local coordinate systems for the primary (M1), the secondary (M2), and the focal plane (FP) are illustrated in orange for the y and z axes (x axes follow the "right-hand rule" convention, they all go into the page). The "world" coordinate system is the intersection of the boresight and elevation axes. The primary instrument space is outlined in blue. The subordinate instrument spaces (black) are located below and closer to the boresight, on either side of the telescope mid-plane.

For a standard Mizuguchi-Dragone telescope, the optical design is completely determined by five parameters, as shown by Granet[19]. Here we specify the diameter of the main reflector and four parameters associated with the central ray, that is, the ray traveling down the boresight axis a shown in Figure 1. These parameters are the deflection angle of the incoming ray off the primary, the distance traveled from the primary and secondary, the distance traveled from the secondary to the focal plan, and the angle between the boresight ray and the ray traveling to the focal plane. Table 1 lists the optical parameters for the initial CCAT-prime/SO LAT design. This results in a focal length of the primary of 28.8 m. The secondary focal length is 18 m and has a conic constant K = -5. The effective focal length of the telescope is $f_{\mathrm{eff}}$ = 14.4 m, which means that the ratio $f_{\mathrm{eff}}$ / $D$m = 2.4.

Table 1. Optical characteristics of the initial design.

| | |
|---|---|
| Aperture Diameter *Dm* | 6 m |
| Offset angle of primary $\theta_0$ | $\tan^{-1}(4/3) = 53.130°$ |
| Distance primary to secondary *Lm* | 6 m |
| Distance secondary to focal plane *Ls* | 12 m |
| Angle between boresight and output $\theta_{\mathrm{p}}$ | 90° |

One of the main requirements of CCAT-prime and SO LAT is the 7.8° field of view (FOV) at 3mm wavelength. (Note that use of a larger FOV is possible with this design; for example, preliminary reimaging optics designs that illuminate 5.5m of the primary mirror and span a 9° FOV have been demonstrated[10].) The classical crossed-Dragone design is known for having a large FOV, but it is still a "classical" design, in that the primary is a paraboloid and the secondary is a hyperboloid. Many on-axis optical telescopes use the Ritchey-Chrétien (RC) geometry to obtain a larger field of view than one would get with a classic Cassegrain design. The RC design corrects for coma aberrations by making small modifications to the shapes of the mirrors. An obvious question to ask is whether the FOV of the crossed-Dragone design can be improved in the same way. We show here that this is indeed possible and that the increase in the area of the diffraction-limited FOV can be large – as much as a factor of 10 at 300 µm wavelength in the case we considered. In fact Dragone[20] pointed out in 1983 that correction of coma should be possible for off-axis telescopes, but as far as we are aware no such telescopes have actually been built. This paper describes the optimization process that we used to make modifications to the mirror surfaces which maximize the diffraction-limited FOV of our telescopes.

In order to implement the modifications to the surfaces in a form that is convenient for both optical analysis and for manufacture, their shapes are described by the "sag" $z$ as a function of position $(x,y)$ in the local coordinate frame centered on the point where the chief ray meets each mirror and aligned with the normal to the surface at that point. We used polynomials to express the sag $z(x,y)$, choosing the coefficients to match the classical shapes in the initial design and then adjusting them in order to obtain a larger field of view in the subsequent optimization. See section 2.3 for details.

### 2.1 Background

In a "classic" two-reflector Cassegrain telescope the primary is a paraboloid and the secondary is a hyperboloid. As one moves off-axis the image quality deteriorates and the principal aberration is usually coma. The Ritchey-Chrétien (RC) modification is to make both reflectors hyperboloids and to tune the choice of their conic constants to cancel the coma. Note that although there is no longer a sharp image at the (imaginary) primary focus, this does not need to compromise the on-axis performance at the secondary focus – the distortion introduced into the wavefront by the primary (mostly a donut-shaped spherical aberration) is removed by the secondary. In general, radio telescopes do not use the RC formulation. The reason for this is that they generally have extremely "fast" primary mirrors – f/D ~ 0.4 – and this means that other aberrations, especially astigmatism, come in very quickly as one goes off-axis, so little can be gained in this way[†]. In our classical Dragone design, by contrast, the optics are relatively "slow" at all stages and this makes it worthwhile to implement modifications to the shapes of the mirrors that correct coma while having only a small impact on other aberrations.

[†] In fact many radio telescopes do use a similar "shaping" of their reflectors in order to improve the illumination – i.e. to make the strongly tapered illumination provided by a typical feed-horn produce a more uniform pattern on the primary aperture. This works in the opposite direction to the RC formulation and actually reduces the field of view.

## 2.2 Coma Correction

Coma aberration in a classical Cassegrain telescope is due to path errors of the form $x^3\theta$, where $x$ is the distance of the ray from the axis at the aperture and $\theta$ is the angle between the object and the telescope axis. (For simplicity we describe just the case where the rays and the offset angle are in the x-z plane.) One way of understanding how the RC formulation works is to realize that the main effect of changing the primary from a parabola to a hyperbola is to introduce a term in $x^4$ into the shape of the primary and that to produce zero wavefront error on-axis there has to be a similar term, with the same magnitude and opposite sign, added to the secondary (by changing its conic constant). As one moves off-axis, the relative positions at which the rays fall on the mirrors changes by an amount which is proportional to $\theta$ and to the distance between the mirrors, $L$. This results in a change in the path length proportional to $\theta L$ times the x-derivative of $x^4$. This means that the coma will be corrected if we choose the right magnitude for the $x^4$ term. When the full 2-dimensional case is considered this becomes a term in $r^4$ for a symmetric telescope.

For an off-axis telescope we would expect that the same principle would apply but the form will need to be somewhat more complicated to allow for the fact that the distance between the mirrors is a function of position in the aperture. In addition, points on a circle in the aperture will fall as ellipses on the mirror surfaces. This means that the modifications to the surfaces will need to contain more than just the simple $r^4$ term required for the RC design. Polynomials in $x$ and $y$, i.e. a sum of terms of the form $x^i y^j$, turn out to provide the necessary generality. Dragone[20] gives an expression for the coefficients of the fourth-order terms but we found that both higher- and lower-order terms are needed and we are not able to derive their coefficients by analysis, so we find them using the optimization procedure in Zemax OpticStudio[21]. We tried two different approaches. The first simply minimizes the rms spot size for a set of points in the focal plane – in practice we chose the center and two rings of points with a strong weight on the center, lower on the first ring and very low on the outer ring. In this case it was also necessary to control the effective focal length and to keep the plate scale equal in the $x$ and $y$ directions. For the second case we made use of the fact that correcting for coma is equivalent to making sure that the optical system matches the Abbe sine condition, which is that, at the center of the FOV, the sine of the angle of arrival of a ray is proportional to its radial distance from the axis in the aperture. A classical Cassegrain antenna violates this condition while an RC satisfies it. We therefore constructed a merit function which explicitly requires that a suitable set of rays should meet this condition. We found that the two approaches gave similar results but that the first produced a slightly larger useable field of view at lower frequencies, probably because it was providing partial correction for some of the other aberrations. The results shown here used the first method.

Figure 2 shows false-color representations of the normal changes in the shapes of the two mirror surfaces that resulted from the optimization – the primary is on the left and the secondary on the right. One sees the expected donut-like deviations with opposite signs, but the donuts are distorted because of the off-axis design. The magnitudes of the deviations are rather small – under one millimeter peak-to-peak.

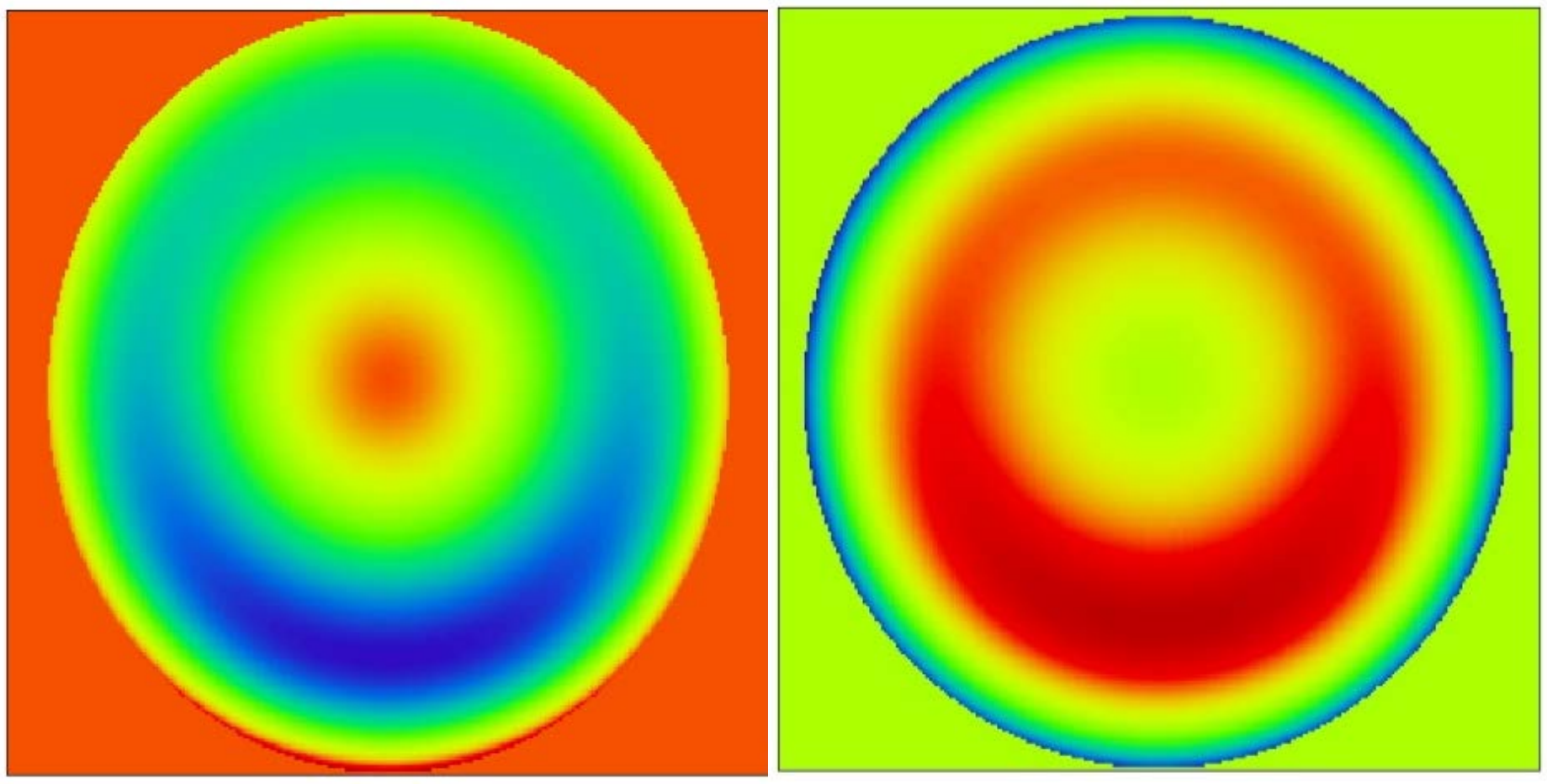

Figure 2. False-color representation of the surface deviation in mm (z plane) from the base conic surface for the primary (left) and secondary (right). The colors red to blue cover the range -0.1 to +0.5mm for the primary and -0.5 to +0.5mm for the secondary.

Figure 3 shows cuts across the mirror surfaces to illustrate better the form and magnitude of the corrections.  Note the horizontal scales are in meters and the vertical scales are in millimeters.

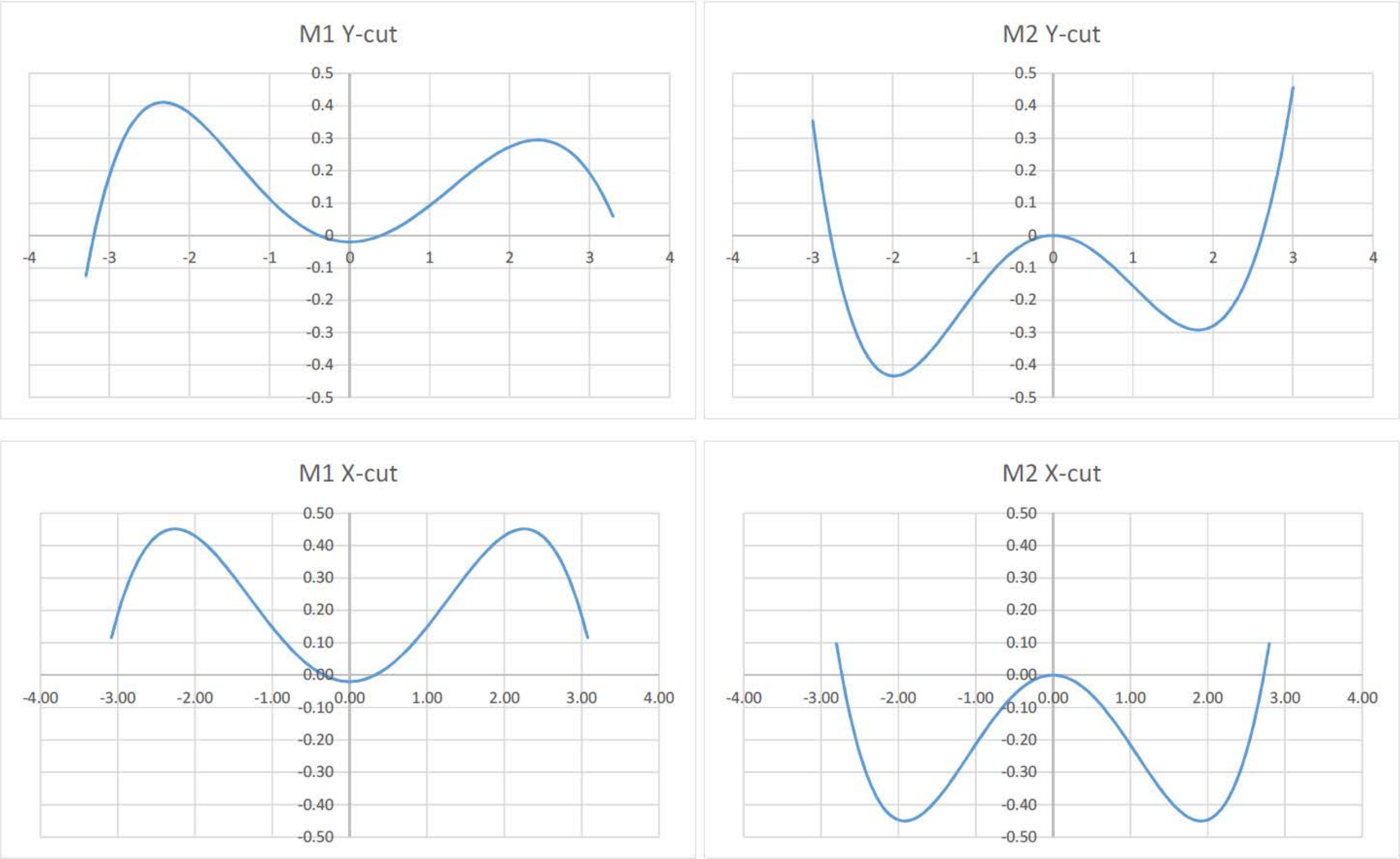

Figure 3.  Cuts across the surfaces are shown for the y and x planes in the upper and lower portions, respectively.  Again this shows only the modification introduced by the optimization.  The horizontal scales are in meters and the vertical scales in mm.

Figure 4 shows the wavefront aberrations as a function of field angle in the +y, +x and –y directions, with the classical design on the left and the coma-corrected design on the right.  It is seen that the coma terms are reduced by more than an order of magnitude.  There is a small increase in the astigmatism and in the higher-order terms but the total wavefront error is reduced significantly, especially at small field angles.  Crucially, the wavefront error for astigmatism is roughly quadratic in field angle, whereas for coma it is essentially linear.  This means that the useable FOV for high frequency observations is significantly increased by correcting the coma, as shown in figure 5.

The useable area has also become more elliptical (Figure 6).  This is because the astigmatism is larger in the Y-direction (the plane perpendicular to the axes about which the mirrors are tilted) than the X-direction.

One side effect of the shaping is to increase the curvature of the field – the radius decreases from ~25 to ~15 meters – and this starts to have an effect in the outer part of the FOV.  In reality this is not a problem as most instrument designs can accommodate this modest amount of curvature.  In fact, two-mirror telescopes of this type will not be telecentric, i.e. in the outer part of the FOV the chief rays arriving at the detectors from the center of the aperture will not be travelling parallel to the optical axis.  Instead they will be coming from a point (the exit pupil) that is about 20 m away in this case.  This means that, if we build the instrument so that the focal surface is curved with a radius of ~17.5 m and have the detectors looking normal to this surface, both the focus and telecentricity will be acceptable.

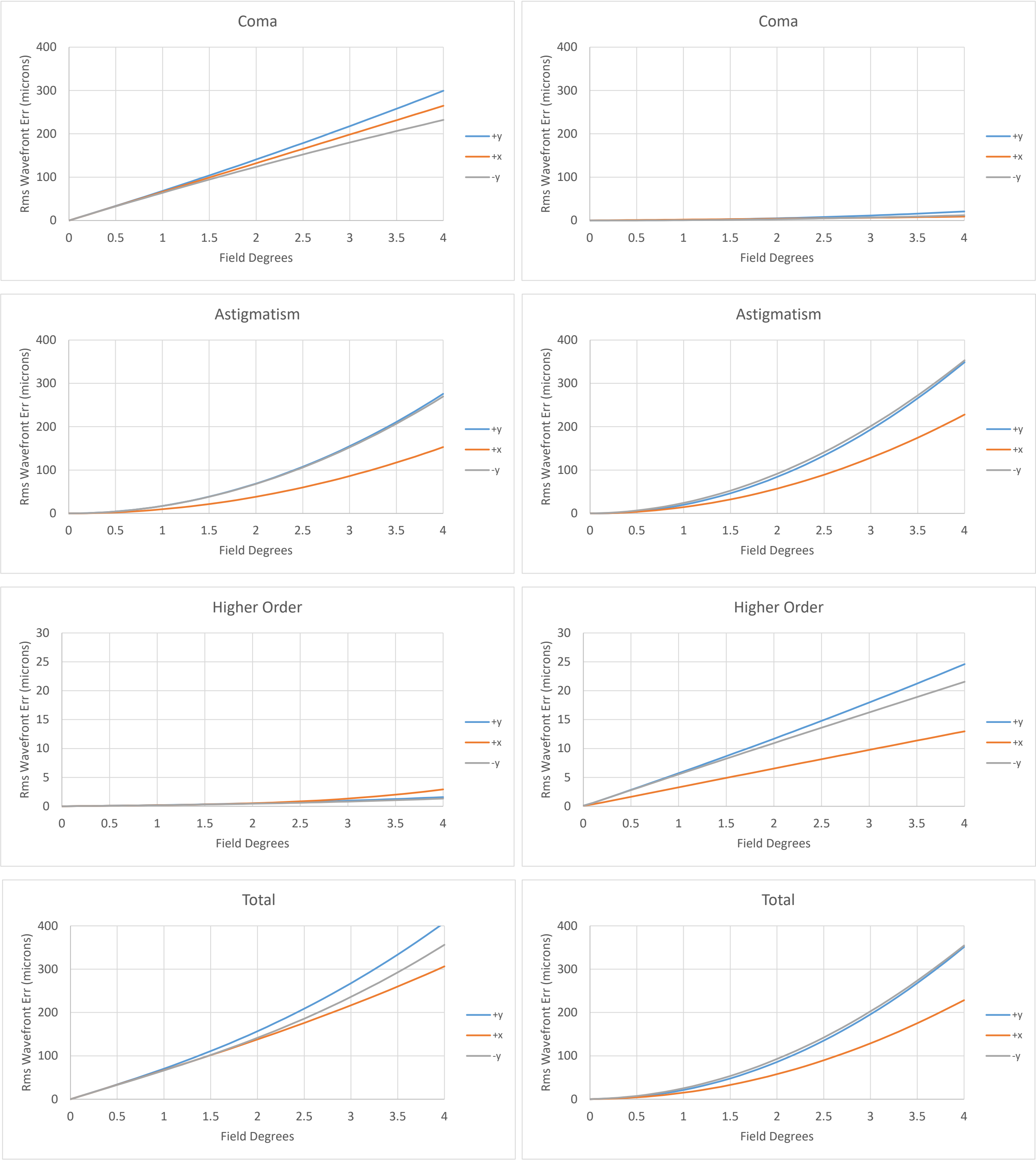

Figure 4. Aberration plots versus field angle for +y, +x and –y directions (blue, red and grey). The classical design is on the left and the corrected design on the right. These are derived from fitting the Standard Zernike polynomials to the wavefront errors. Note that there are two coma terms (with cos and sine dependence on azimuth) and two astigmatism terms. The plots show the rss values of the two terms in each case. The plots for higher order terms show the rss values for all the remaining terms up to 7$^{th}$ order.

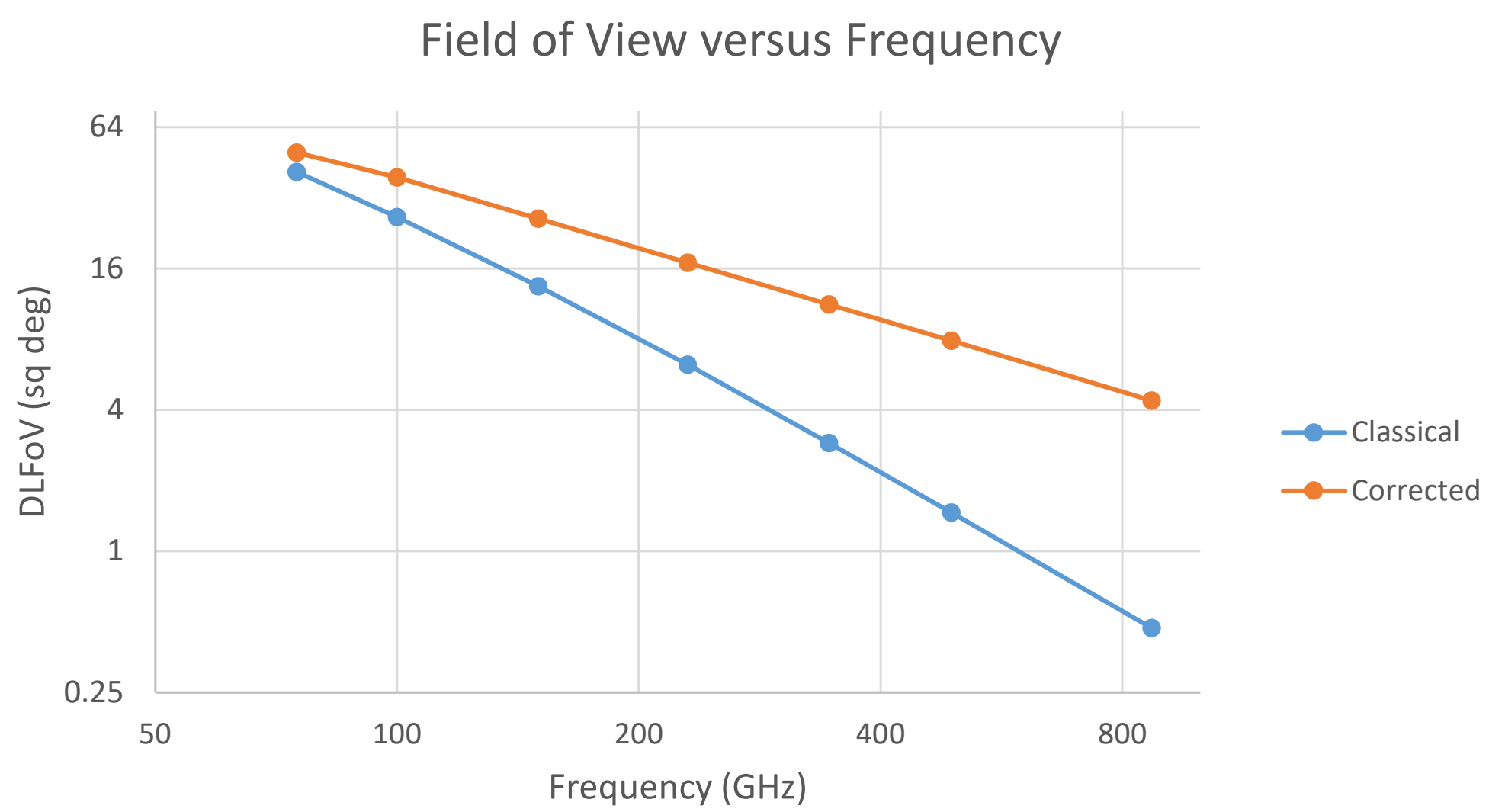

Figure 5. Diffraction limited FOV as a function of frequency. This is the area in the focal plane (within the range +/- 4 deg in each coordinate) within which the Strehl ratio is greater than 80%. This is for the full 6m diameter aperture and with no apodization (i.e. uniform illumination): using a smaller aperture or a tapered illumination would increase the FOV further. Note however that this accounts only for the wavefront errors due to the optics – manufacturing and alignment errors are not included.

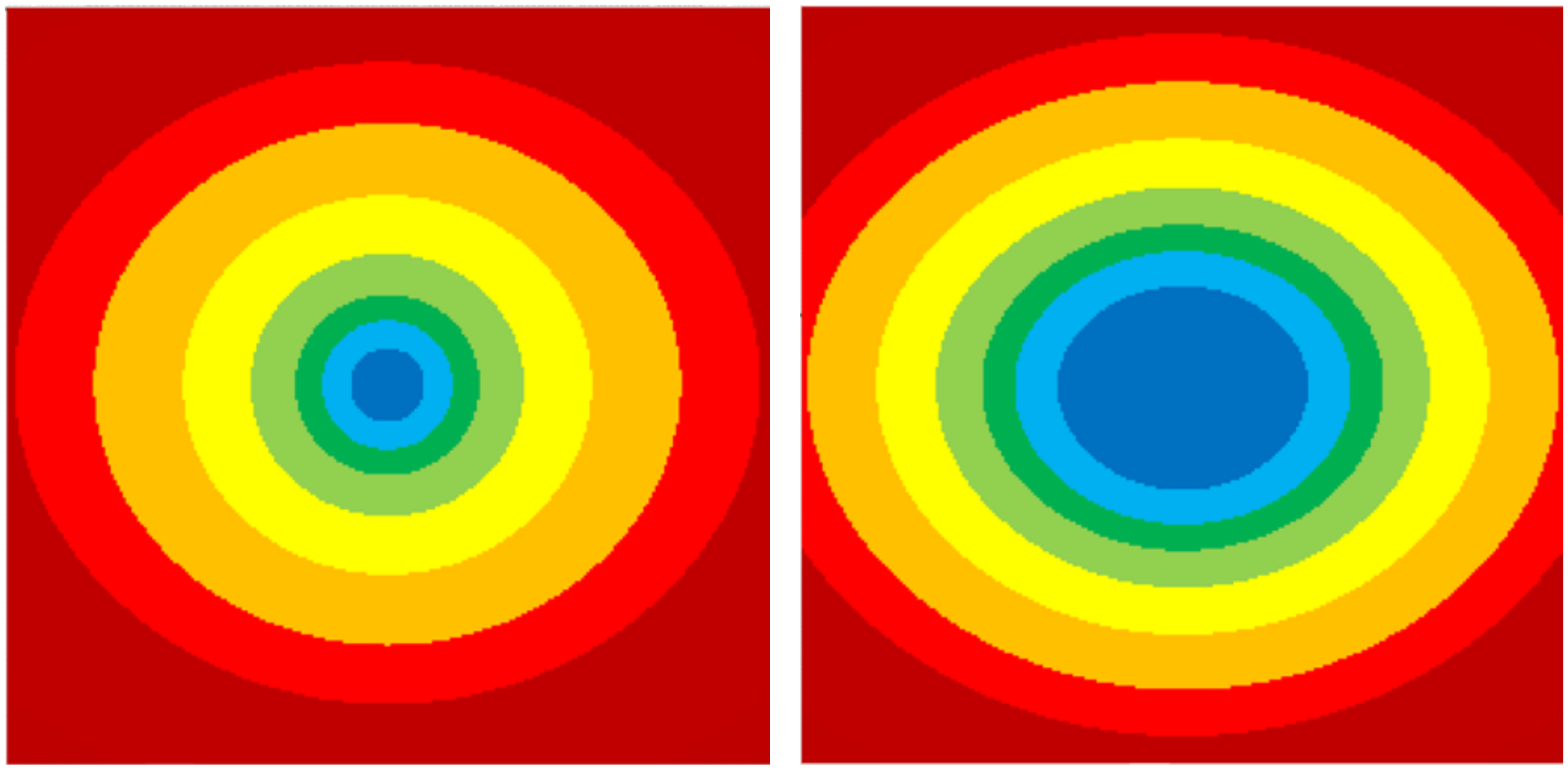

Figure 6. Diffraction-limited regions in an 8 by 8 degree field as a function of frequency with the standard Dragone design on the left and the coma-corrected optics on the right. The colors (blue through red) show the regions where the Strehl ratio is > 80% at 870, 490, 345, 230, 150, 100 and 75 GHz respectively.

Given that one obvious application of this very large field of view would be for observations of CMB polarization, an investigation of the polarization properties of the modified design was conducted in GRASP[22] and is discussed in Gallardo et al. 2018. It was found that the shaping maintains the very low cross polarization intrinsic to the Dragone design.

Characteristics of the final design are given in Table 2. The linear dimensions are rounded to the nearest millimeter. The diffraction limited field of view (DLFOV) is taken at 80% Strehl ratio for full primary illumination and given in degrees for three different wavelengths (300 μm, 1 mm, and 3 mm). As described in Dicker[10], instrument reimaging optics can often be used to achieve diffraction-limited image quality beyond the DLFOV.

Table 2. Characteristics of the production design optics.

| | Production Design Optics | |
|---|---|---|
| Mirror | Primary (M1) | Secondary (M2) |
| center offset from local origin, Y direction (mm) | +35 | -1 |
| major axis (mm) | 6708 | 6167 |
| minor axis (mm) | 6000 | 5844 |
| max sag (mm) | 70 | 112 |
| surface area ($m^2$) | 31.6 | 28.3 |
| Working f/# | 2.6 | |
| Plate scale | ~14 arcsec / mm | |
| DLFOV @ 0.3 mm, X – Y (deg) | 2.46 x 1.94 | |
| DLFOV @ 1.0 mm, X – Y (deg) | 4.54 x 3.65 | |
| DLFOV @ 3.0 mm, X – Y (deg) | 7.91 x 6.34 | |

### 2.3 Surface definition

As explained, the mirror surfaces are defined by a polynomial equation. The general form we used is:

$$z(x,y) = \sum_{i=0}^{k} \sum_{j=0}^{k} a_{ij} \left(\frac{x}{R_N}\right)^i \left(\frac{y}{R_N}\right)^j$$

where $z$ is the mirror sag at a given $x$ and $y$ position, all relative to the local coordinate system of the mirror, $R_N$ is the normalization radius (a scaling factor having no bearing on the domain of the formula), $k$ is the maximum polynomial power, and $a_{ij}$ are the coefficients. For the primary $k$ = 6 and for the secondary $k$ = 7, and both use $R_N$ = 3000 mm. Coefficients for the primary and secondary surfaces are given in Tables 3 and 4, respectively.

Table 3. Coefficients ($a_{ij}$) for the equation for the primary mirror surface.

| $a_{ij}$ | $j$=0 | $j$=1 | $j$=2 | $j$=3 | $j$=4 | $j$=5 | $j$=6 |
|---|---|---|---|---|---|---|---|
| $i$=0 | 0 | 0 | -57.74022 | 1.5373825 | 1.154294 | -0.441762 | 0.0906601 |
| $i$=1 | 0 | 0 | 0 | 0 | 0 | 0 | 0 |
| $i$=2 | -72.17349 | 1.8691899 | 2.8859421 | -1.026471 | 0.2610568 | 0 | 0 |
| $i$=3 | 0 | 0 | 0 | 0 | 0 | 0 | 0 |
| $i$=4 | 1.8083973 | -0.603195 | 0.2177414 | 0 | 0 | 0 | 0 |
| $i$=5 | 0 | 0 | 0 | 0 | 0 | 0 | 0 |
| $i$=6 | 0.0394559 | 0 | 0 | 0 | 0 | 0 | 0 |

Table 4. Coefficients ($a_{ij}$) for the equation for the secondary mirror surface.

| $a_{ij}$ | $j$=0 | $j$=1 | $j$=2 | $j$=3 | $j$=4 | $j$=5 | $j$=6 | $j$=7 |
|---|---|---|---|---|---|---|---|---|
| $i$=0 | 0 | 0 | 103.90461 | 6.6513025 | 2.8405781 | -0.7819705 | -0.0400483 | 0.0896645 |
| $i$=1 | 0 | 0 | 0 | 0 | 0 | 0 | 0 | 0 |
| $i$=2 | 115.44758 | 7.3024355 | 5.7640389 | -1.578144 | -0.0354326 | 0.2781226 | 0 | 0 |
| $i$=3 | 0 | 0 | 0 | 0 | 0 | 0 | 0 | 0 |
| $i$=4 | 2.9130983 | -0.8104051 | -0.0185283 | 0.2626023 | 0 | 0 | 0 | 0 |
| $i$=5 | 0 | 0 | 0 | 0 | 0 | 0 | 0 | 0 |
| $i$=6 | -0.0250794 | 0.0709672 | 0 | 0 | 0 | 0 | 0 | 0 |
| $i$=7 | 0 | 0 | 0 | 0 | 0 | 0 | 0 | 0 |

One question of interest is how many terms are needed to describe the surfaces to sufficient accuracy. Figure 7 shows the rms residuals over the surface as a function of the order of the polynomial used. This suggests that 6$^{th}$ order is nearly good enough to not significantly impact the CCAT-prime surface accuracy (the errors are below 1 µm per surface). Given the goal that CCAT-prime has for surface errors (7.1 µm rms) we extend the definition for M2 to 7$^{th}$ order.

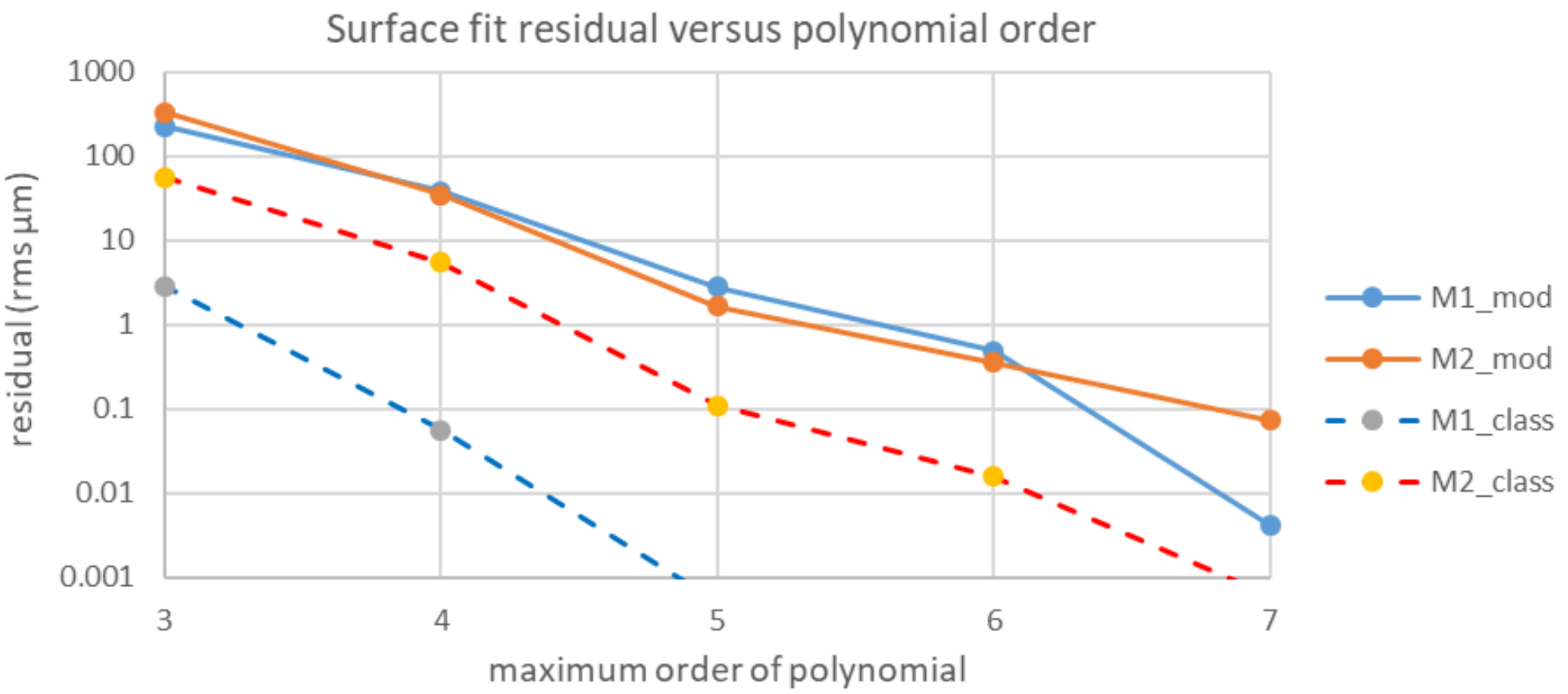

Figure 7. This plot shows the residuals in the modified (M1_mod, M2_mod) and classic (M1_class, M2_class) surfaces, solid and dashed lines respectively. For CCAT-prime, less than 1 µm would be adequate, so 4$^{th}$ and 5$^{th}$ order polynomials would be sufficient for M1 and M2 in describing the classic cross-Dragone system. As expected, the modified design requires a few more terms. We chose 6$^{th}$ and 7$^{th}$ order for M1 and M2 in modeling the coma corrected version.

## 2.4 Layout

Although it is not particularly difficult to find designs that meet the Mizuguchi-Dragone criterion for symmetrical illumination (resulting in low cross-polarization and wide field of view), the resulting layout normally has dimensions that are not rational numbers. However, there is a design which has the appealing property that all the important dimensions have integer values (see Figure 8). In this design, the path traced by the chief ray inside the telescope forms a triangle with sides in the proportion 3:4:5 and other lengths in the optical system are whole numbers related to these. This makes for an elegant and exact description of the optical system, a useful feature when working across various computer programs (e.g., ray tracing, EM modeling, and mechanical design) and multiple collaborations and working groups.

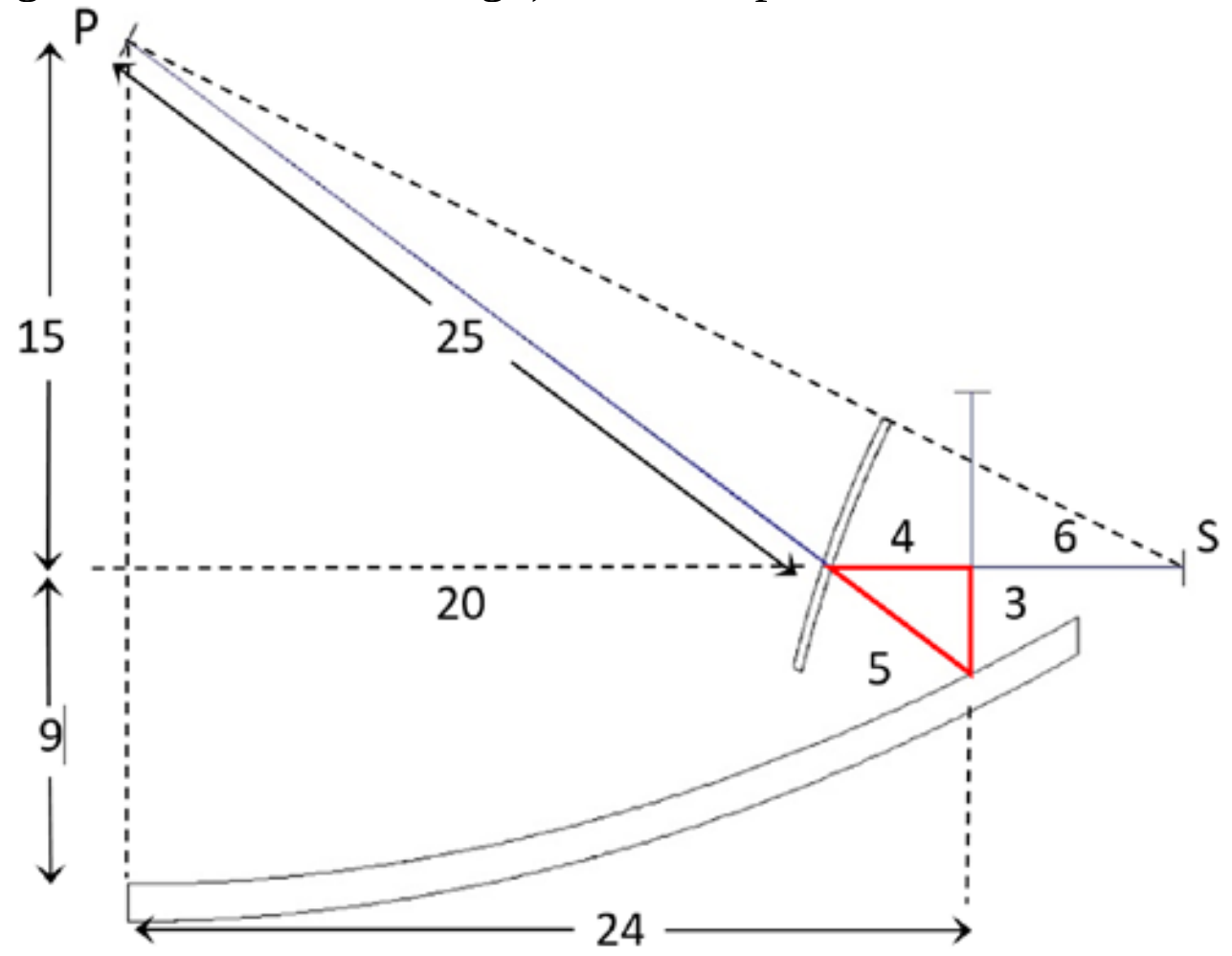

Figure 8. This diagram shows (half of) the parabolic primary (M1) and the hyperbolic secondary (M2 – again half) along with prime and secondary foci P and S. The 3:4:5 triangle is outlined in red and the other lengths are given in the same arbitrary units. To get dimensions suitable for CCAT-prime and SO LAT these values are multiplied by 1200 mm.

# 3 TOLERANCING

Analysis of the rigid body optical tolerances for the modified crossed-Dragone optics is critical in order to optimize the design of the telescope structure. The coefficients for one half wavefront error (HWFE) and pointing errors (PE) in elevation (EL) and cross-elevation (XEL) are determined. To avoid aperture effects and better reflect the proposed instrument designs, the primary was illuminated with a 5500 mm diameter beam (f/2.6) for the sensitivity study. For tolerancing, we focus on CCAT-prime which requires tighter performance specifications for its higher frequency operation.

## 3.1 HWFE sensitivity

For sources at the field center and edges, Zemax OpticStudio (ZOS) was used to determine Strehl ratio degradation versus position error for the secondary (M2) and the focal plane (FP). Each element was investigated for its six degrees of freedom: three translations and three rotations. The primary (M1) and FP were fixed while the position and orientation of M2 was adjusted, and likewise for the FP, M1 and M2 were fixed, while the FP was adjusted.

The Strehl ratio degradation model, and the Strehl ratio / HWFE relation are (respectively):

$$S = e^{-C\left(\frac{\delta}{\lambda}\right)^2}, S = e^{-\left(4\pi\frac{HWFE}{\lambda}\right)^2}$$

where $S$ is the Strehl ratio, $C$ is a constant, $\delta$ is a perturbation (translation or rotation), and $\lambda$ is the wavelength. Thus, fitting the data from the results of the optical model provides constants that convert position error (in µm or arcsec) to HWFE in rms µm, i.e. we are interested in HWFE / $\delta$ (see Table 5). As an additional check, the rms wave error from ZOS was recorded, and when converted to HWFE in rms µm, agreed with the values derived from the Strehl ratio degradation method. Wavefront gradient is not included in the Strehl ratio degradation results.

## 3.2 Pointing error sensitivity

Pointing errors for M2, the FP, and bulk motion about the local origin of M1 for M1, M2 and FP as a solid unit, were determined using the ZOS model, measuring shifts in the focal plane, and converting to EL and XEL terms using the plate scale of the telescope. Figure 9 shows how far-field and focal plane coordinates are mapped to telescope pointing errors of EL and XEL.

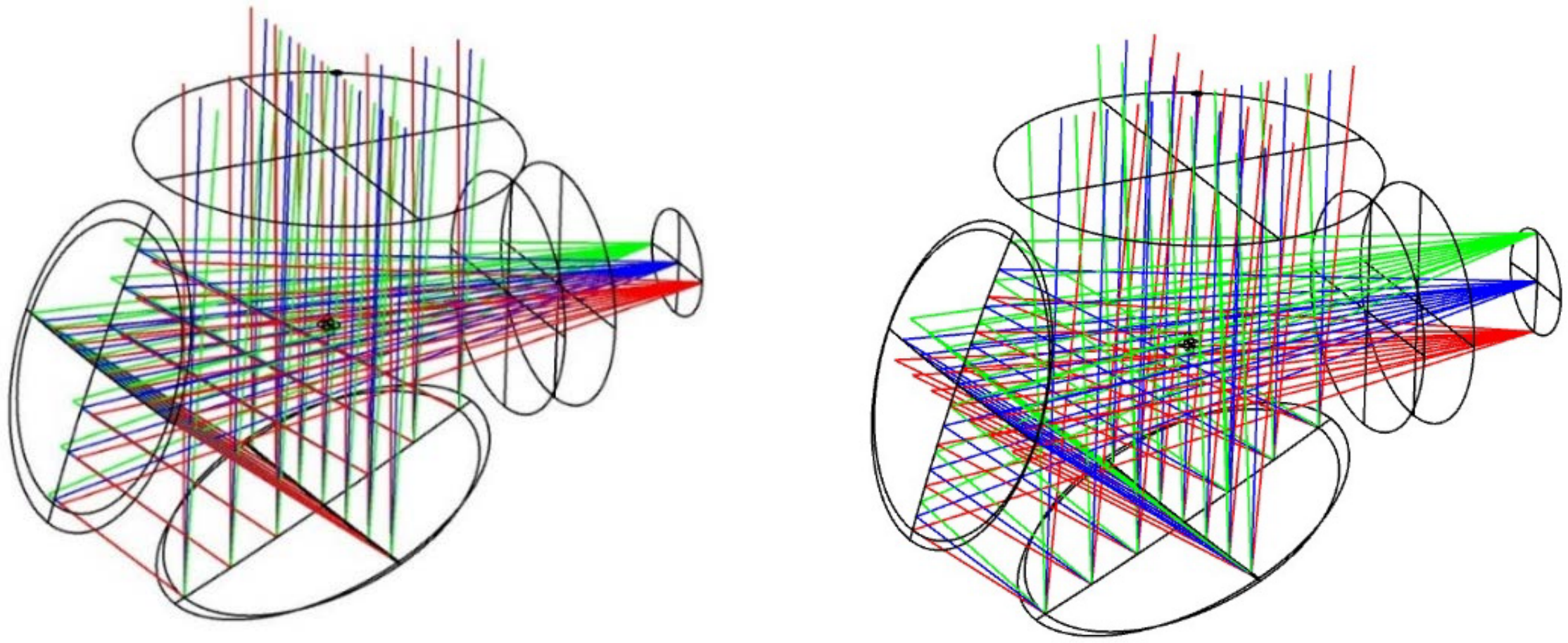

Figure 9. Diagrams defining how ZOS field / focal plane X and Y coordinates translate to telescope elevation and cross-elevation terms. Left: the +X (green), center (blue) and -X (red) fields & focal plane coordinates. By convention, EL is measured up from the horizon, with M2 on the left, the focal plane on the right, and one is looking at the horizon from the telescope. A +EL error corresponds to a -X shift in the FP. Right: the +Y (green), center (blue), and -Y (red) fields and focal plane coordinates. XEL is measured positive right of the EL plane of motion. Using the same orientation, M2 on the left, FP on the right, a +XEL error corresponds to a +Y shift in the FP.

### 3.3 Results

Table 5 lists the sensitivity coefficients for HWFE, PE-EL, and PE-XEL due to motions of the optical elements. The entire telescope was translated and tilted in each direction (x, y, z, α, β, γ) with respect to the local coordinate system of M1 (see Figure 1). This has no effect on HWFE since all the elements move together. For PE, translations have no effect on far field pointing (i.e., all science observations), leaving only rotations, α, β, and γ, about the x, y, and z axes, respectively. The M1 x-axis is parallel to the XEL axis and so of course any α errors map 1 to 1. The β and γ coefficients map to pointing errors in EL as given in the table. It is worth noting that these values are equal to cos(atan(1/2)) and sin(atan(1/2)) – the orientation of M1 – for β and γ, respectively. This provides a convenient check on the validity of the ZOS model.

For the secondary and its local coordinate system (Figure 1), rotating γ has insignificant effect on the pointing error for the center field, but does impact the pointing for off-axis field positions, although only to a minor degree. It is noted that limiting γ to +/- 700 arcsec keeps the PE for the edges of the field to a negligible level, where negligible is taken as 1% of the size of a 3.1 mm wavelength beam. The M2 α and β rotations show the highest sensitivity. Decenter (x, y) and despace (z) all have non-zero values for HWFE and PE-EL or PE-XEL.

Decentering (x, y) the focal plane produces a pointing error according to the plate scale, but has no impact on the HWFE. From the ZOS model, despace (i.e., defocus – z) has an impact on the HWFE but no effect on pointing. Focal plane α, β, and γ tilts have no effect on HWFE or PE for the center field. Limiting α and β tilts to +/- 600 arcsec and limiting γ to +/- 20 arcsec results in negligible PE for the edges of the field. Additionally, α and β tilts will shift the illumination of the primary by $\delta_{shift}$ {mm} = 0.070*$\theta_{FP}$ {arcsec}. A margin of 10 mm shift can be allocated in the design so the α and β tilts must be limited to about 140 arcsec, equivalent to a tip of about 1.4 mm across the focal plane diameter.

Table 5. Sensitivity results. HWFE is in rms µm per µm-displacement (x, y, z) or rms µm per arcsec-rotation (α, β, γ). PE is in arcsec per µm-displacement (x, y, z) or arcsec per arcsec-rotation (α, β, γ).

| Telescope Optics | Motion | HWFE | PE-EL | PE-XEL | Notes |
|---|---|---|---|---|---|
| Bulk Motion (about M1) | | | | | |
| | decenter, x | n/a | 0 | 0 | no HWFE; no far field PE |
| | decenter, y | n/a | 0 | 0 | no HWFE; no far field PE |
| | despace, z | n/a | 0 | 0 | no HWFE; no far field PE |
| | tilt, α | n/a | 0 | 1.000 | no HWFE |
| | tilt, β | n/a | 0.894 | 0 | no HWFE |
| | tilt, γ | n/a | 0.447 | 0 | no HWFE |
| Secondary (M2) | | | | | |
| | decenter, x | 0.00034 | -0.0083 | 0 | |
| | decenter, y | 0.00076 | 0 | 0.0079 | |
| | despace, z | 0.0029 | 0 | -0.0064 | |
| | tilt, α | 0.044 | 0 | -1.66 | |
| | tilt, β | 0.042 | -1.57 | 0 | |
| | tilt, γ | 0.014 | 0 | 0 | no PE for center field, negligible PE for field edges if -700" < γ < 700" |
| Focal Plane (FP) | | | | | |
| | decenter, x | 0 | 0.014 | 0 | negligible HWFE for shifts < 20 mm |
| | decenter, y | 0 | 0 | -0.014 | negligible HWFE for shifts < 20 mm |
| | despace, z | 0.0026 | 0 | 0 | no PE |
| | tilt, α | 0 | 0 | 0 | no HWFE or PE for center field; require -140" < α < 140" to limit M1 illumination shifts to < 10 mm; negligible HWFE & PE for field edges if -600" < α < 600" |
| | tilt, β | 0 | 0 | 0 | no HWFE or PE for center field; require -140" < β < 140" to limit M1 illumination shifts to < 10 mm; negligible HWFE & PE for field edges if -600" < β < 600" |
| | tilt, γ | 0 | 0 | 0 | no HWFE or PE for center field; negligible HWFE & PE for field edges if -20" < γ < 20" |

### 3.4 Discussion

It is clear from Table 5 that the tolerances are much more relaxed for HWFE than for PE. However, repeatable PE can be taken out with a systematic pointing error model (SPEM), so the real concern is the random and/or uncorrectable PE and any residual PE from the SPEM. The coefficients for the bulk motion of the telescope with respect to the M1 origin might prove useful in canceling some of the other pointing error terms.

The HWFE for translations requires positional tolerances of a few 100's µm worst case for < 1 µm HWFE degradation. Limiting the HWFE to < 1 µm for the α, β and γ tilts of M2 is also a challenge, requiring the mirror skew to be constrained to the order of 1 mm across its surface. Such positioning could likely be achieved with a very carefully designed steel structure.

Regarding PE, allocating one half of the offset pointing error budget for M2 α or β rotation (the most sensitive coefficients) requires the mirror skew to be < 13 µm across the face of the mirror. This would be very difficult to achieve in a steel structure of this size assuming typical temperature gradients, but reasonable using carbon fiber composites or Invar. The focal plane requires careful attention to changes in x, y, and γ, with 10's of µm (arcsec) potentially having a non-negligible effect.

## 4 CONCLUSION

Although a classic Dragone optical design delivers a large useable FOV, applying a RC modification to the design leads to substantial improvements, especially at shorter wavelengths. The design for CCAT-prime and SO LAT provides a relatively flat focal plane to support large format cameras. With a field of view of 7.8° in diameter at 3 mm wavelength, and the ability to illuminate >100k diffraction-limited beams of < 1 mm wavelength, both telescopes represent an order of magnitude improvement over current facilities.

The optics consist of just two mirrors, a primary and secondary, arranged in such a way as to satisfy the Mizuguchi-Dragone criterion, which suppresses first-order astigmatism and maintains high polarization purity. The mirror surface shapes are modified from standard conics in order to suppress first-order coma.

The tolerancing coefficients show the design is highly sensitive to orientation (tip-tilt) of M2, but within reasonable limits of materials and fabrication techniques. HWFE is also most sensitive to the M2 orientation. Keeping the pointing error low means the HWFE due to misalignment is minimized. Limits have been found for all FP rotations based on having negligible impact to the pointing of a 3.1 mm beam, a reasonable approach since only the longest wavelengths will be used at the field edges.

These two telescopes will provide the largest diffraction-limited optical throughput at millimeter and sub-millimeter wavelengths of any observatories yet. They will serve as important pathfinders for next generation microwave observatories, such as the proposed CMB-S4 project[23].

**Acknowledgements:** CCAT-prime funding has been provided by Cornell University, the Fred M. Young Jr. Charitable Fund, the German Research Foundation (DFG) through grant number INST 216/733-1 FUGG, the Univ. of Cologne, the Univ. of Bonn, and the Canadian Atacama Telescope Consortium. Part of this work was supported by a grant from the Simons Foundation (Award #457687, B.K.). MDN acknowledges support from NSF award AST-1454881. RD thanks CONICYT for grants PIA Anillo ACT-1417 and QUIMAL 160009.